\documentclass{article} 
\usepackage[final]{colm2025_conference}

\usepackage{microtype}
\usepackage{hyperref}
\usepackage{url}
\usepackage{booktabs}
\usepackage{graphicx}
\usepackage{subcaption}
\usepackage{longtable}

\usepackage{lineno}

\definecolor{darkblue}{rgb}{0, 0, 0.5}
\hypersetup{colorlinks=true, citecolor=darkblue, linkcolor=darkblue, urlcolor=darkblue}

\title{More Parameters Than Populations: A Systematic Literature Review of Large Language Models within Survey Research}

\author{Trent D Buskirk\\
School of Data Science\\
Old Dominion University\\
Norfolk, VA 23529, USA \\
\texttt{tbuskirk@odu.edu} \\
\And
Florian Keusch \\
School of Social Sciences \\
University of Mannheim \\
Mannheim, Germany \\
\texttt{f.keusch@uni-mannheim.de} \\
\And
Leah von der Heyde \\
Department of Statistics \\
LMU Munich \\
Munich, Germany \\
\texttt{l.heyde@lmu.de} \\
\And
Adam Eck \\
Computer Science Department \\
Oberlin College \\
Oberlin, OH 44074, USA \\
\texttt{aeck@oberlin.edu} \\
}

\begin{document}

\ifcolmsubmission
\linenumbers
\fi

\maketitle

\begin{abstract}
Survey research has a long-standing history of being a human-powered field, but one that embraces various technologies for the collection, processing, and analysis of various behavioral, political, and social outcomes of interest, among others. At the same time, Large Language Models (LLMs) bring new technological challenges and prerequisites in order to fully harness their potential. In this paper, we report work-in-progress on a systematic literature review based on keyword searches from multiple large-scale databases as well as citation networks that assesses how LLMs are currently being applied within the survey research process. We synthesize and organize our findings according to the survey research process to include examples of LLM usage  across three broad phases: pre-data collection, data collection, and post-data collection. We discuss selected examples of potential use cases for LLMs as well as its pitfalls based on examples from existing literature. Considering survey research has rich experience and history regarding data quality, we discuss some opportunities and describe future outlooks for survey research to contribute to the continued development and refinement of LLMs.
\end{abstract}

\section{Introduction}

Large language models (LLMs) are rapidly transforming many professional domains, including survey research. \cite{eloundou2024} rank survey research among the most highly exposed occupations to LLM-driven automation, raising both opportunities and challenges for practitioners. While survey research has a rich tradition of adopting technological tools for tasks like data collection, analysis, and instrument design \citep[e.g.,][]{waksberg1978,baker1992,couper2000,couper2013}, the unique affordances and risks associated with LLMs call for a structured examination. \cite{jansen2023} provide largely a conceptual overview of the potential uses and considerations for incorporating LLMs within the survey research context. Since their work was published, the field is transitioning from an ideation phase to an implementation phase. This paper aims at expanding our understanding of the potential and concerns of this technology within the survey research context by presenting work-in-progress on a systematic literature review of empirical and theoretical work at the intersection of LLMs and survey research. Specifically, we synthesize examples of how LLMs are being applied across three broad phases of the survey research pipeline: pre-data collection, data collection, and post-data collection.

\section{Systematic Literature Review}

\begin{figure}[htbp]
    \centering
    \includegraphics[width=\linewidth]{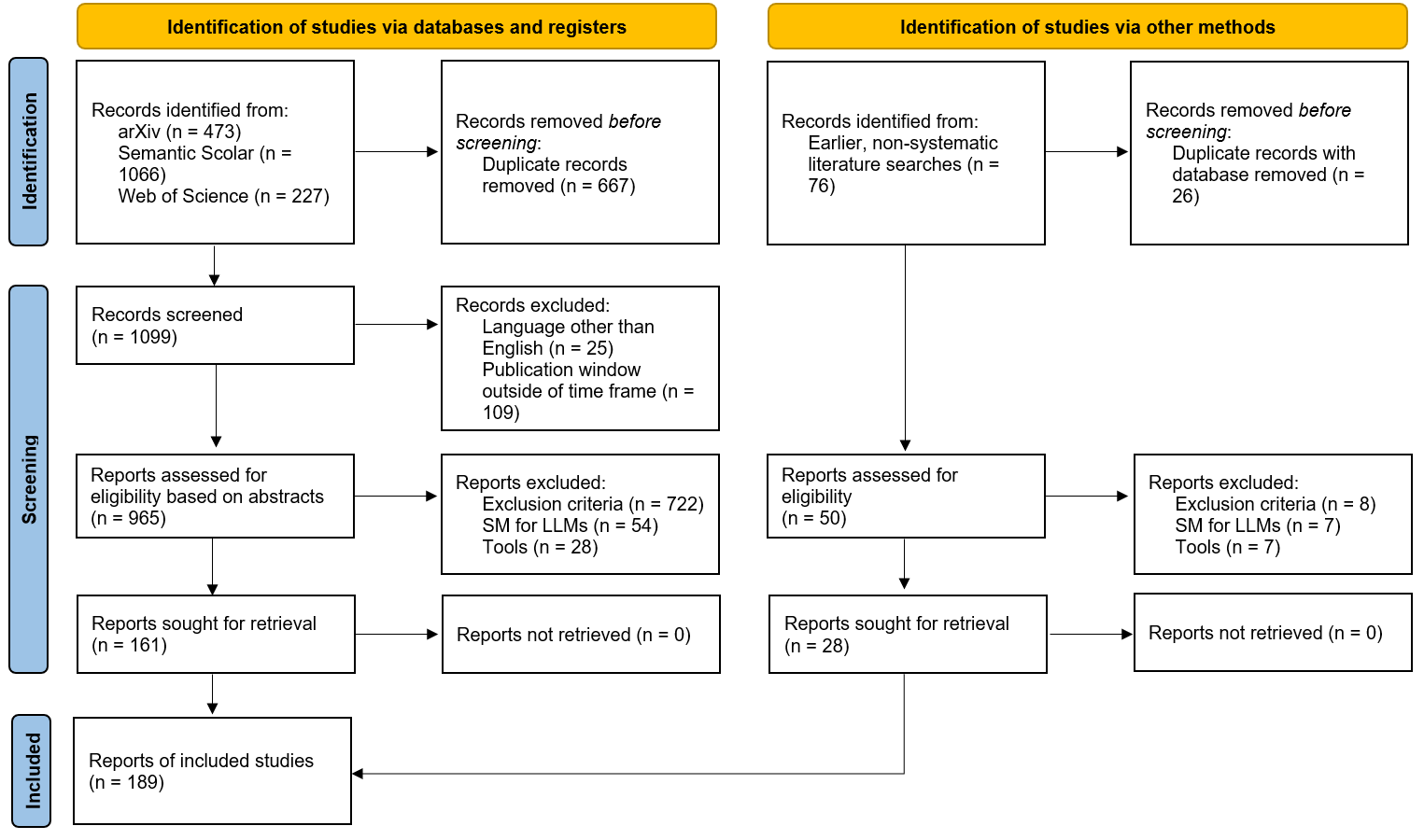}
    \caption{PRISMA 2020 flow diagram for systematic review \citep{page_prisma_2021}}
    \label{fig:fig1}
\end{figure}

Our systematic literature review process identified 1,766 candidate publications retrieved from keyword searches using 10 LLM- and 62 survey-related keywords across arXiv, Semantic Scholar, and Web of Science databases and citation networks (see Figure~\ref{fig:fig1}). After deduplication across these three databases, we retained 1,099 unique papers. Of those, 134 were not written in English or appeared to be outside of our 6-year publication window spanning from January 2019 through January 2025 and were thus excluded. The remaining 965 papers underwent abstract review by members of our team and produced a total of 161 papers that were deemed eligible for inclusion (see Inclusion and Exclusion Criteria in the Appendix). In addition, we identified another set of 28 papers  as citing or cited papers not already included in the collection of 161 eligible papers. One of the authors performed a review of the total of 189 papers that discuss applications of LLMs within the survey research process. \\

\section{Application of LLMs in Three Phases of the Survey Process}

We classified the papers in our corpus based on in which of three phases of the survey process an LLM is applied. The \textit{pre-data collection phase} considers tasks such as questionnaire writing, item generation, translation, sampling designs, and recruitment materials and efforts. The \textit{data collection phase} encompasses AI-assisted interviews, silicon sampling, and similar tasks that result in produced data. Finally, the \textit{post-data collection phase} includes data processing tasks, weighting, imputation, summarization, report generation, and dataset curation.

\begin{figure}[htbp]
    \centering
    \begin{subfigure}{0.48\linewidth}
        \centering
        \includegraphics[width=\linewidth]{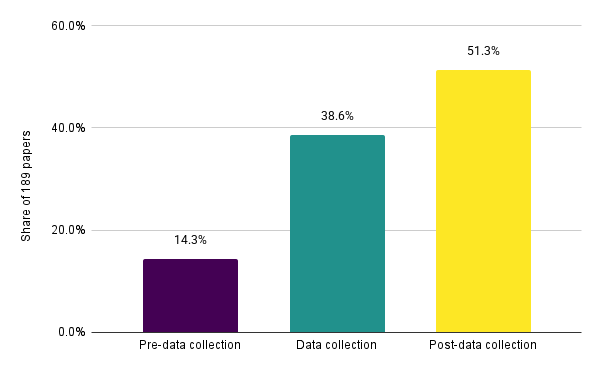}
        \caption{Distribution by phase}
        \label{fig:fig2a}
    \end{subfigure}
    \hfill
    \begin{subfigure}{0.48\linewidth}
        \centering
        \includegraphics[width=\linewidth]{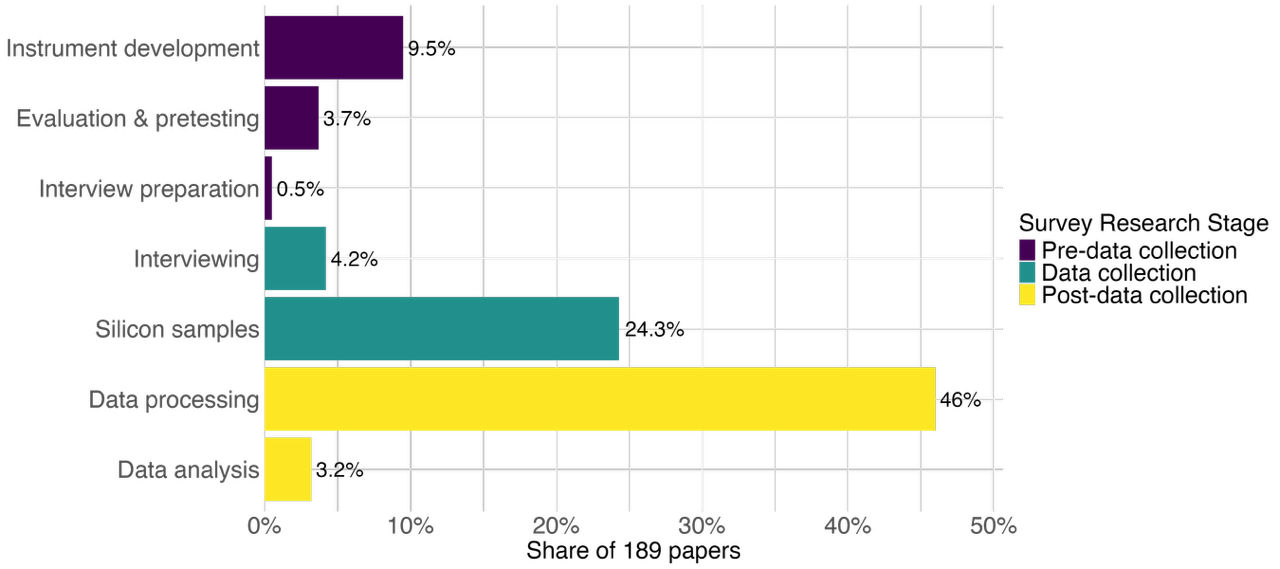}
        \caption{Distribution by application}
        \label{fig:fig2b}
    \end{subfigure}
    \caption{Distribution of papers by phase and application}
    \label{fig:fig2}
\end{figure}

Initial findings indicate that empirical applications of LLMs concentrate in three areas (see Figure~\ref{fig:fig2}): instrument development, synthetic respondent modeling, and automated text classification. For instance, \cite{adhikari2025} demonstrate how LLMs can simulate cognitive interviews, enabling automated pretesting of cross-cultural questionnaires. Their method adapts survey items from U.S. contexts to new populations (e.g., South Africa), revealing both the promise and pitfalls of LLM personas in usability assessment. For generating synthetic data, \cite{cho2024} introduce LLM-based “doppelgänger” models, which fine-tune LLMs on individual-level conversational data to forecast participant responses across a survey, raising questions about generalizability, personalization, and bias. LLMs have also been deployed for open-text data classification, such as the work of \cite{li2024} on accessibility sentiment extracted from Google Maps reviews to inform urban design. Meanwhile, \cite{bisbee2024} and \cite{mellon2024} issue cautionary notes, highlighting the risk of LLMs hallucinating responses or misclassifying nuanced political opinion data—thereby raising concerns about representational accuracy and construct validity in “silicon samples.”\\

\section{Potential New Tools}

We also saw a smaller number of papers that we identified as tools or methods that are being developed with LLMs that have the potential to be applied in the survey research process but haven’t yet been. For example, work is being done to identify LLM output \citep{nathanson2024} that could be used to assist in fraudulent respondent detection. There are also new methods and frameworks being developed for synthetic samples and data generation using LLMs \citep[e.g.,][]{balog2024} that could enhance related synthetic sampling approaches that are currently being explored within survey research.

\section{Survey Research Improving LLMs}

While we focused on papers in which LLMs were used within the survey research process, there were a fair number of excluded papers that pointed to ways in which survey research could also improve LLMs. A few themes that emerged among these papers were the curation and creation of training or benchmark data for improving or evaluating LLM output \citep{xie2024} as well as the use of surveys to understand perceptions and use cases of LLMs in various industries \citep{gasparini2024, johnson2024}.

\section{Conclusion}

Methodologically, the current state of our review identifies an uneven distribution of LLM applications across the survey pipeline. While pre-data collection stages (e.g., item writing, translation) are well explored, core practices such as live interviewing, recruitment, and cross-lingual adaptation remain under-investigated. This stands in contrast to LLMs’ key features as multi-modal, multi-lingual processors and producers of natural language, offering room for further important research. Additionally, few studies assess LLMs systematically across multiple populations, languages, or survey topics, leaving questions about the generalizability of successful applications.

Our work highlights not just the breadth of current use cases, but also the methodological and ethical considerations that must accompany them. Given survey research’s long-standing expertise in assessing error sources and ensuring data quality, it is well-positioned to guide responsible and effective LLM integration into scientific workflows.

\section{Next Steps}

In a next step, we will continue the work on the review of the 189 papers in our corpus. We will implement double-coding by augmenting the manual coding performed by one of the authors with an AI-based approach. In addition to coding in what specific phase of the survey process and for what particular application the LLMs are used, we will also classify whether papers are theoretical or empirical, which LLM(s) are discussed, which populations are covered or languages are applied, the substantive domain and topic, and under what conditions the application is deemed successful. Going forward, our review will reveal how LLM training data, model architecture, as well as research designs \citep[see, e.g.,][]{vonderHeyde2025} pose challenges to augmenting survey research with LLMs.

\section*{Author Contributions}
\textbf{Trent Buskirk:} Conceptualization, Methodology, Writing - Original Draft. \textbf{Florian Keusch:} Conceptualization, Methodology, Writing - Review \& Editing. \textbf{Leah von der Heyde:} Conceptualization, Methodology, Formal Analysis, Investigation, Data Curation, Writing - Review \& Editing, Visualization. \textbf{Adam Eck:} Conceptualization, Methodology, Investigation, Data Curation, Writing - Review \& Editing.

\section*{Acknowledgments}
We would like to acknowledge our student research assistants who assisted with our literature review: Elinor Frost and Reefayat Bin Shahjahan at Oberlin College and Ella Häcker and Leonard Bek at the University of Mannheim.

\nocite{*}

\bibliography{BKHE_NLPOR2025}
\bibliographystyle{colm2025_conference}

\newpage

\appendix
\section{Appendix}
\section*{Inclusion criteria}

\begin{itemize}
    \item \textbf{Screening}
    \begin{itemize}
        \item \textbf{English} language
        \item \textbf{Published between 2019 and before search date} (Feb 4, 2025)
    \end{itemize}

    \item \textbf{Eligibility}
    \begin{itemize}
        \item Access to \textbf{full paper}
        \item Paper about use of \textbf{LLMs in survey research or public opinion research}
        \item \textbf{Theoretical discussion} or \textbf{empirical application}
        \item Use of LLMs in...
        \begin{itemize}
            \item \textbf{pre-data collection or generation} including \underline{questionnaire development} (e.g., item generation, translation), \underline{evaluation}, \underline{pretesting} (e.g., cognitive interviewing, focus groups), \underline{sample design} (e.g., frame construction, sampling plan), \underline{recruitment preparation} (e.g., drafting of invitation letters, interviewer training), and \underline{keyword selection};
            \item \textbf{data collection or generation}, including \underline{sample management} (e.g., call scheduling), \underline{data generation} (e.g., silicon samples), and \underline{AI interviewing} (LLMs embedded for augmentation or used for autonomous interviewing);
            \item \textbf{post-data collection or generation}, including \underline{data processing} (e.g., the extraction or coding of open-ended questions, social media data, or other digital trace data indicative of public opinion or behavior, such as comments on news articles; weighting, imputation); \underline{data analysis}, \underline{summarization}, and \underline{visualization}; and \underline{dissemination} (e.g., report writing, codebook creation, creation of public use files)
        \end{itemize}
        \item Data that represents \textbf{information about human attitudes and behaviors} to include survey data, administrative data, social media, and digital trace data (e.g., user posted comments or blog posts) and sensor data.
        \item \textbf{LLMs} include everything starting with BERT (i.e., ``pre-trained transformers'')
    \end{itemize}
\end{itemize}

\section*{Exclusion criteria}

\begin{itemize}
    \item \textbf{Screening}
    \begin{itemize}
        \item \textbf{Non-English} language
        \item \textbf{Published before 2019 or after search date} (Feb 4, 2025)
    \end{itemize}

    \item \textbf{Eligibility}
    \begin{itemize}
        \item \textbf{Only abstract} available
        \item Papers about \textbf{``surveys'' of LLM literature} (i.e., literature reviews)
        \item Papers about \textbf{opinions/attitudes about LLMs/AI or their use}
        \item Papers using LLMs in the context of \textbf{(Q\&A) testing/evaluation and diagnostics} (e.g., education, medicine, law)
        \item Papers \textbf{conducting lab or field experiments} with LLMs as the unit of analysis [Survey Research helping LLMs category]
        \item Papers exclusively using \textbf{news content} 
        \item Papers that develop \textbf{methods that are not used in context of surveys and public opinion research yet} (but could be in the future) [TOOL category]
        \item Papers that \textbf{use survey data to assess quality of LLMs} [SM helping LLMs category]
        \item Papers using ML/NLP methods that are \textbf{not LLMs or predate BERT LLMs}
    \end{itemize}
\end{itemize}

\end{document}